# Thermal Conductivity of Double Polymorph Ga$_2$O$_3$ Structures

Azat Abdullaev[1,2*], Kairolla Sekerbayev[1,2], Alexander Azarov[3], Vishnukanthan Venkatachalapathy[3], Vinay S. Chauhan[4], Zhandos Utegulov[2*], Andrej Kuznetsov[3*]

[1]*Center for Energy and Advanced Materials Science, National Laboratory of Astana, Nazarbayev University, Kabanbay Batyr avenue 53, 010000 Astana, Kazakhstan*

[2]*Department of Physics, School of Sciences and Humanities, Nazarbayev University, Kabanbay Batyr avenue 53, 010000 Astana, Kazakhstan*

[3]*Department of Physics and Centre for Materials Science and Nanotechnology, University of Oslo, N-0316 Oslo, Norway*

[4]*Intel Corporation, Chandler, 85226, Arizona, USA*

**Abstract**

Recently discovered double gamma/beta (γ/β) polymorph Ga$_2$O$_3$ structures constitute a class of novel materials providing an option to modulate functional properties across interfaces without changing chemical compositions of materials, in contrast to that in conventional heterostructures. In this work, for the first time, we investigate thermal transport in such homo-interface structures as an example of their physical properties. Specifically, the cross-plane thermal conductivity ($k$) was measured by femtosecond laser-based time-domain thermoreflectance with MHz modulation rates, effectively obtaining depth profiles of the thermal conductivity across the γ/β-Ga$_2$O$_3$ structures. In this way, the thermal conductivity of γ-Ga$_2$O$_3$ $k$=1.84÷2.11 W m$^{-1}$K$^{-1}$ was found to be independent of the initial β-substrates orientations, in accordance with the cubic spinel structure of the γ-phase and consistently with the molecular dynamics simulation data. In its turn, the thermal conductivity of monoclinic β-Ga$_2$O$_3$ showed a distinct anisotropy, with values ranging from 10 W m$^{-1}$K$^{-1}$ for [201] to 20 Wm$^{-1}$K$^{-1}$ for [010] orientations. Thus, for double γ/β Ga$_2$O$_3$ polymorph structures formed on [010] $\beta$-substrates, there is an order of magnitude difference in thermal conductivity across the $\gamma/\beta$ interface, which potentially can be exploited in thermal energy conversion applications.

I. Introduction

Fabrication of heterostructures and functionalization of their properties is among the most successful strategies in solid-state technology, specifically for designing new electronic, magnetic and thermal properties in semiconductors. There is a conventional way to fabricate such heterostructures by changing chemical compositions of materials across interfaces, typically realized in-situ during thin film synthesis[1]. Thus, there are numerous examples of highly useful heterostructures between chemically dissimilar semiconductors with tunable thermal properties[2–5]. On the other hand, utilizations of stacks between different phases in chemically identical materials, i.e. across homo-interfaces, are rare, perhaps with an exception of the crystalline/amorphous silicon structure, exploited in high-efficiency solar cells[6]. Nevertheless, functional properties of the crystalline-to-crystalline polymorph junctions have not been investigated in a systematic way. This may be explained by a limited availability of such samples having both reasonable quality and technological relevance.

Indeed, there is practically no literature data reporting double polymorph structures synthesis with conventional deposition methods having technologically sufficient quality of interfaces between polymorphs. This is quite understandable because the temperature/pressure growth conditions are strongly different for various polymorphs and must be changed abruptly, likely leading to the interfacial quality degradation. Concurrently, even though polymorph structures may be induced by the localized pressure application to already existing crystals, the control over the resulting structures is limited by the accuracy/scalability of the pressure application instrumentation[7]. On the other hand, a new method to fabricate double $\gamma/\beta$-$Ga_2O_3$ polymorph structures by a self-organized $\gamma$-polymorph transformation upon reaching a certain disorder threshold induced into $\beta$-$Ga_2O_3$ by irradiation was recently reported in literature[8–12]. Importantly, the formation of such double polymorph $\gamma/\beta$-$Ga_2O_3$ structures is featured by a paradoxically good sharpness of the $\gamma/\beta$ interface[10], making such structures comparable with conventional heterostructures in terms of the functionalization options. As such, stacking different crystal symmetry polymorphs, in particular cubic $\gamma$-phase and monoclinic $\beta$-phase of $Ga_2O_3$, may result into the step-like changes of thermal properties across these homo-interfaces, potentially exploitable in phononic devices in line with that of heterointerfaces between chemically dissimilar materials[13–17].

Importantly, even though $Ga_2O_3$ attracts a lot of attention as a promising ultra-wide bandgap semiconductor, the understanding of its thermal transport properties is immature; e.g. experimental data are documented only for its stable $β$-$Ga_2O_3$ polymorph[18–20]. On the other hand, the thermal conductivity ($k$) data for the metastable $Ga_2O_3$ polymorphs are rare and limited to theoretical predictions, e.g., for the $α$-phase[21] and $ε$-phase[22]. This may be attributed to the issues with the growth of stable polymorphs of reasonable quality. Thus, in this work, for the first time, we investigated heat conduction properties of the double polymorph $γ/β$-$Ga_2O_3$ structures, fabricated by a self-organized $γ$-polymorph transformation induced in $β$-$Ga_2O_3$ by irradiation[9,10,12]. For this reason, we used femtosecond laser-based time-domain thermoreflectance (TDTR) measurements with MHz modulation frequency to investigate heat conduction properties on a set of the double $γ/β$ polymorph $Ga_2O_3$ structures with variable top $γ$-layer thicknesses (ranging over hundreds of nanometers) fabricated on the $β$-$Ga_2O_3$ substrates having different crystallographic orientations. By comparing the experimentally measured data with the results of molecular dynamic (MD) simulations, we demonstrate up to an order of magnitude difference in thermal conductivities between $γ$- and $β$-phases depending on the crystalline orientation, providing additional options for thermal functionalization of the $Ga_2O_3$ devices.

## II. Experimental Section/Methods

### A. Materials

As initial samples, we used high-purity pristine β-$Ga_2O_3$ substrates with three different crystal orientations such as [010], [201] and [100], purchased from Novel Crystal Technology Inc. All samples have had lateral size of 5×5 mm and were polished from the side exposed to the ion irradiation. In order to obtain $γ$-$Ga_2O_3$, all $β$-$Ga_2O_3$ samples were irradiated with $5×10^{16}$ Ga/cm$^2$ ions which is known to be above the disorder threshold needed for converting the top part of the $β$-phase to $γ$-phase at room temperature[12]. Notably, until recently, there was a puzzle with the identification of the new phase formed in the course of the disorder-induced ordering of $β$-$Ga_2O_3$. Some of the earlier works attributed the newly emerging polymorph to the orthorhombic κ-phase[9,23], however now it is unambiguously identified that $β$-phase converts into the cubic spinel, i.e. the γ-phase[8,10,12,24]. Importantly, in order to vary the thickness of the top γ-phase layer, we used three different Ga ion energies, literally, 0.5, 1, and 1.7 MeV, at constant fluence of 5×10$^{16}$ Ga/cm$^2$.

*B. Structural characterization*

To control the structural properties of the double Ga$_2$O$_3$ γ/β polymorph, we used a combination of the x-ray diffraction (XRD) and Rutherford backscattering spectroscopy in channeling mode (RBS-C), relating this study to systematic structural characterization by extensive transmission electron microscopy imaging[8–12,23,24]. The XRD 2-theta scans were performed using Bruker AXS D8 Discover diffractometer applying Cu Kα1 radiation in locked-coupled mode. The RBS-C analysis was carried out using 1.6 MeV He$^+$ ions, and was also used to measure the thickness of the newly formed top γ-layers. It is important to note that *β-Ga$_2$O$_3$ is not amorphized under irradiation, but upon reaching a certain disorder threshold, transforms into γ-phase* as shown previously[9,10,12,23], with a characteristic box-like shape of the RBS-C spectra used as fingerprints of the double Ga$_2$O$_3$ γ/β polymorph in this study.

*C. Thermal transport measurements and simulations*

For the thermal transport measurements, we used a frequency-modulated (TDTR) setup, see details elsewhere [25,26]. Briefly, Ti:sapphire mode-locked femtosecond laser (Tsunami, Spectra-Physics) at 782 nm wavelength, having 80 MHz repetition rate and 80 fs pulse duration, was used as the pump and the probe beams. The pump beam, modulated by an electro-optic modulator over 0.73-10 MHz thermally excites the samples and control heat penetration depth by $D_{th} = \sqrt{\frac{k}{\pi C_v f}}$, where $k$ is the thermal conductivity, $C_v$ is the volumetric specific heat capacity and $f$ is the pump modulation frequency. The probe beam is optically scan-delayed by a motorized delay stage and then detected by a photodetector connected to a radio-frequency lock-in amplifier. Both beams are focused on the sample surface by a 10x objective lens resulting in the 1/e$^2$ diameter of ~15 μm which is substantially larger than the heat diffusion length. Therefore, TDTR measurements provide sensitivity only to cross-plane $k$ at this beam diameter. A heat diffusion model was used to extract $k$ and the interface thermal boundary conductance ($G$) of the studied samples by fitting a time-delayed signal ratio of the in-phase to out-of-phase voltages (–V$_{in}$/V$_{out}$) on a picosecond scale [27]. Notably, Al heat transducing layer we deposited on all samples to enable the TDTR measurements. The thickness of this transducer was determined by the transient picosecond acoustic measurement and cross-checked with profilometer data, while the thermal conductivity of the Al transducer layer was measured using a reference sample, see Supplementary Materials.

To model $k$, we implemented classical molecular dynamics (MD) simulations using both the equilibrium (EMD) and non-equilibrium (NEMD) molecular dynamics. For the EMD simulations, the lattice $k$ was modeled with the Green-Kubo formalism relating the conductivity to the heat current autocorrelation function, see details in Supplementary Materials. NEMD modeling was also performed in comparison with our measured and EMD outcomes. We used reverse NEMD calculations[27] where a heat source was set in the center of the simulation box with two heat sinks placed at opposing edges. The details are given in Supplementary Materials.

### III.   Results and Discussion

Figure 1 shows the summary of the experimental data for the phase-transformed $Ga_2O_3$ including (a) schematics of the sample preparation, (b-c) examples of the structural analysis, and (d) representative thermal conductivity data. Fig. 1 (b) demonstrates the XRD data of the samples fabricated on [010] oriented $β$-$Ga_2O_3$. There is a broad peak at ~ 63.7°, in accordance with literature identified as (440) cubic spinel $γ$-phase plane reflection forming as a result of the disorder-induced transition [8–12,23,24]. The increase of the $γ$-phase peak intensity along with the increase of the $Ga^+$ ion energy is attributed to the broadening of the $γ$-phase film on top of bulk $β$-$Ga_2O_3$ substrate as illustrated in the schematics in Fig. 1 (a) and as systematically confirmed in literature [8–12]. For two other $β$-$Ga_2O_3$ crystal orientations, the evolutions of the $β$-to-$γ$-phase transition were similar to that in Figs. 1 (a-c) (not shown); also consistently with earlier observations[9,10,28].

Figure 1 (c) shows RBS-C spectra of the [010]-oriented $β$-$Ga_2O_3$ samples irradiated with different energies, in comparison with the channeling and random data for the unimplanted sample. The top axis shows the corresponding depth scale. Importantly, we observe a tilted baseline for the unimplanted sample (as a signature of the normal dechanneling) and the "box-like" RBS-C signal on the top of the tilted baseline for all irradiated samples as a result of the radiation disorder induced $β$-to-$γ$-phase transition[9,10,28]. Apparent broadening of the "box-like" part of the spectra as a function of the $Ga^+$ ion energy is the evidence of the $γ$-phase layer thickening . The estimated $γ$-layer thickness in 0.5 MeV irradiated samples was ~ 350 nm, while for 1 MeV and 1.7 MeV the $γ$-layer thickness proliferates to ~ 650 nm and 1000 nm, respectively. This thickness variation is indicated by vertical dashed lines in Fig. 1 (c). Notably, none of the samples reached the "random"

RBS-C level confirming that β-to-γ-phase transformation *is not accompanied by the amorphization*, consistently with literature [8–10,23,28]. Heat conduction measurements also confirmed this RBS-C observation of not reaching "amorphous limit", as discussed below.

The *greatest novelty* of this work was in measuring the thermal conductivity in the phase-transformed structures as illustrated in Fig. 1 (d); however before discussing these data we have to introduce the TDTR measurements on pristine *β*-Ga$_2$O$_3$, for which we used a two-layer (Al/*β*-Ga$_2$O$_3$) model. Basic properties of *β*-Ga$_2$O$_3$ such as specific heat and density were taken from literature[18–20]. For reference, examples of the multi-frequency fitting to assess $k$ values of β-Ga$_2$O$_3$ for different crystal orientations are available in Supplementary Materials. At this end, Fig. 1 (d) plots the *effective thermal conductivity* ($k_{eff}$) as a function of $f$ in the sample irradiated with 0.5 MeV, i.e. through the depth of the *γ*/*β*-Ga$_2$O$_3$ structure. It is worth to note that $k_{eff}$ was obtained assuming a two-layer model (Al/Ga$_2$O$_3$), which includes fitting the conductivity of the whole Ga$_2$O$_3$ layer and thermal interface conductance with metal transducer layer. A non-uniform $k_{eff}$-trend is observed in Fig. 1 (d) as a function of the heat penetration depth ($D_{th}$), nicely correlating with the double polymorph *γ*/*β* nature of the sample as illustrated by the sample cartoon in the inset in Fig. 1(d). Indeed, $k_{eff}$ is constant within the *γ*-layer and starts to increase at ~1.9 MHz corresponding to $D_{th}$ ~ 338 nm. Importantly, this depth is consistent with 350 nm *γ*-layer thickness as determined by RBS-C data in Fig. 1(c). Therefore, TDTR measurements also confirm the formation of a new layer with a lower thermal conductivity on the top of the *β*-Ga$_2$O$_3$ substrate. To assess $k$ of γ-phase, we perform thermal data analysis based on a three-layer model (Al/γ-Ga$_2$O$_3$/β-Ga$_2$O$_3$). A similar analysis has been previously used for swift heavy ion irradiated single crystalline sapphire, where we used the same approach to spatially resolve the thermal conductivity of radiation-induced subsurface amorphous Al$_2$O$_3$ layer and the ion track region of irradiated Al$_2$O$_3$[25]. Al and *β*-Ga$_2$O$_3$ properties were known from measurements of the unimplanted samples and were fixed during the fitting procedure for the samples containing γ-layers. Two unknown adjustable parameters, $k$ of γ-Ga$_2$O$_3$ and conductance $G$ between Al and γ-Ga$_2$O$_3$ phase were obtained using the multi-frequency fitting (see Supplementary Materials). Importantly, in accordance with literature, the interface between γ and β-phases is remarkably abrupt[8,9,24]. Accounting that the sensitivity analysis showed minor variations of $G$ between γ and β-phases (see Supplementary Materials), in the current analysis $G$ was set to a fixed value while the thickness of the γ/β interface was assumed to be ~8 nm, as could be conservatively estimated from literature[9,12].

The parameters used for fitting are summarized in Table SII in the Supplementary Materials. Thus, for γ-phase we determined $k = 1.84 \pm 0.1$ W m$^{-1}$ K$^{-1}$ in the sample with the thinnest γ-film. This value is an order of magnitude lower than that in the bulk [010] oriented β-phase. However, it is larger than the thermal conductivity corresponding to the amorphous gallium oxide, estimated from Cahill's minimum limit[29,30] ~ 1.15 W m$^{-1}$ K$^{-1}$.

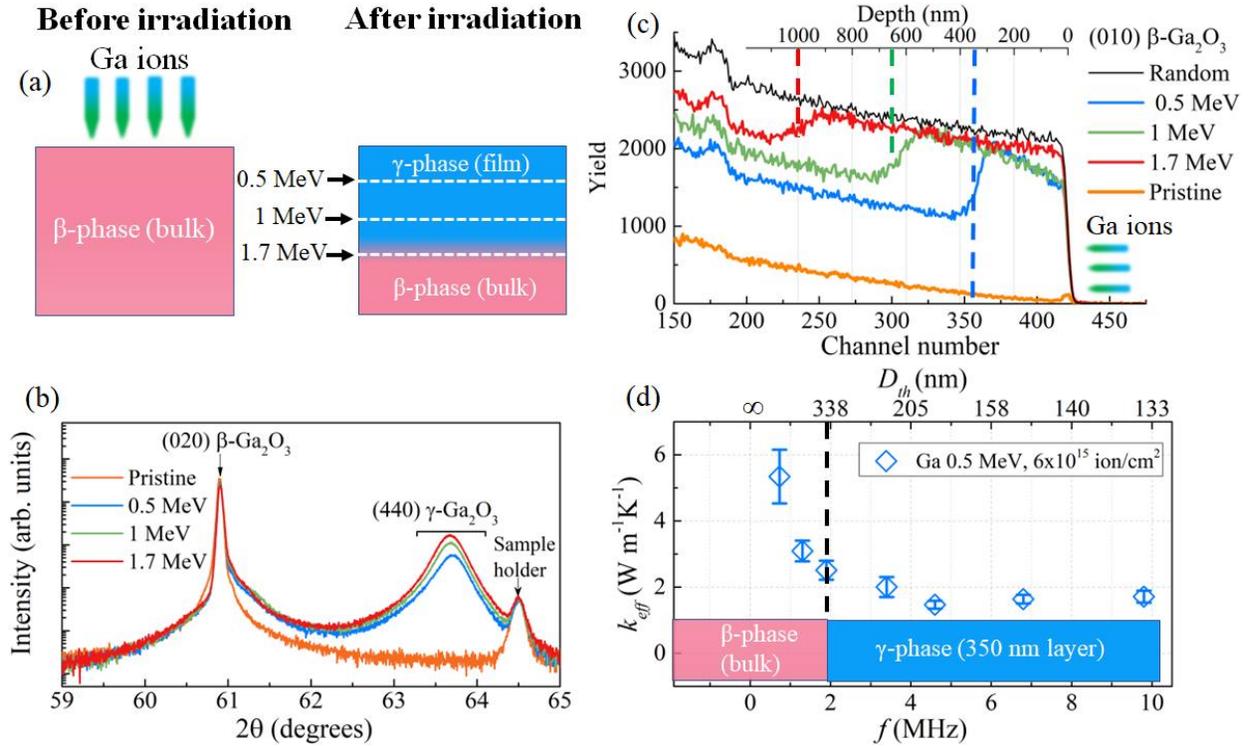

Fig.1 (a) Schematics of the double γ/β-Ga$_2$O$_3$ polymorph structure formation with variable thicknesses of the γ-film depending on the ion irradiation energy, as well as corresponding (b) 2 theta XRD scans and (c) RBS-C spectra. Notably, the dashed lines in panel (c) indicate the positions of the γ/β interfaces. Panel (d) plots the effective thermal conductivity ($k_{eff}$) as a function of the TDTR pump frequency ($f$) - also converted into the heat penetration depth $D_{th}$ at the upper axis - for the sample irradiated with 0.5 MeV Ga ions. The inset into panel (d) is a schematic of the sample cross-section. All data were collected on the samples fabricated with [010] oriented β-substrates.

Figure 2 plots $k_{eff}$ data as a function of the frequency for variable (a) Ga$^+$ ion energies and (b) crystal orientations. Importantly, in Fig. 2 (a) we observe no $k_{eff}$-variations in the samples irradiated by 1 and 1.7 MeV ions, as opposed to that in 0.5 MeV irradiated sample. This is readily explainable by the γ-phase thickening to ~650 and ~1000 nm for samples irradiated with 1 MeV and 1.7 MeV ions, respectively, as illustrated by the RBS-C data correlating with $D_{th}$ not exceeding 500 nm at the lowest modulation rate of 0.73 MHz. It implies that with TDTR at this modulation frequency we measure $k_{eff}$ exclusively of the γ-phase, but not of the underlying β-phase. The results of the multi-layer fitting applied with the 350, 650 and 1000 nm thick γ-films, corresponding to the irradiations with 0.5 MeV, 1 MeV and 1.7 MeV ions respectively, are summarized in Table I. Notably, for all previously studied semiconductors, one would expect a drop in $k$ with the increase of ion energy because of increased defect production. However, the results in Fig. 2 (a) show rather constant trend, perhaps exhibiting a slight increase in $k$ as a function of the irradiation energy, which can be attributed to even better stabilization of the γ-phase obtained via disorder-induced ordering in the higher energy irradiated samples.

Further, Fig. 2 (b) confirms that $k$ of the newly formed γ-phase is independent of the crystal orientation and remains at ~ 1.8 W m$^{-1}$ K$^{-1}$. These results indicate that the phase transformation in β-Ga$_2$O$_3$ occurs independent of the crystal orientation, consistently with literature[9,28]. Another important correlation is that $k$ of the cubic spinel γ-phase is isotropic, in contrast the β-phase anisotropy, as also illustrated in Fig. 2 (b) by comparing $k$ values for different crystallographic directions in β-Ga$_2$O$_3$.

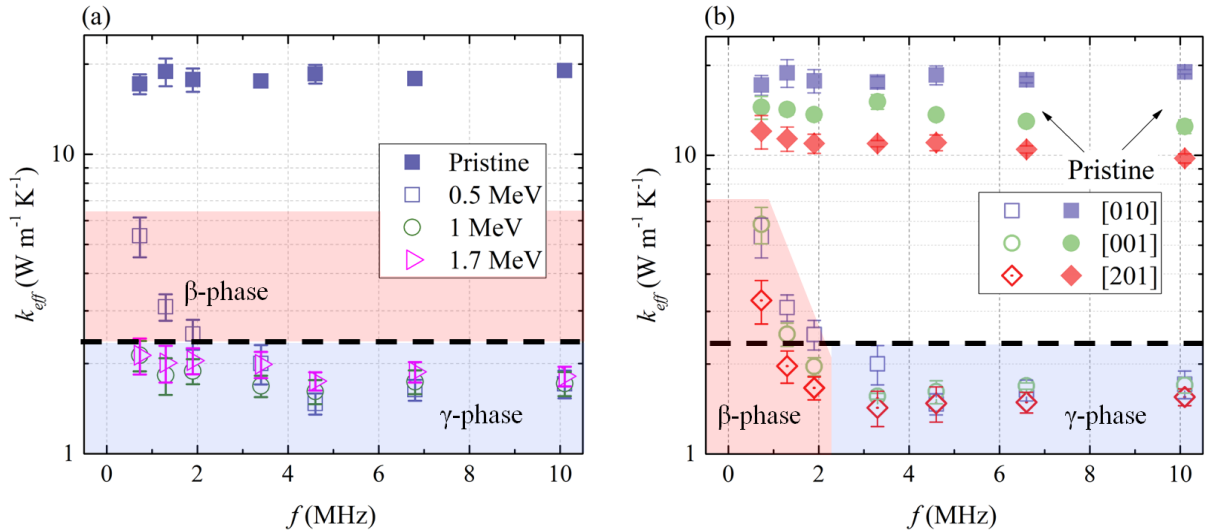

Fig. 2. Modulation frequency-dependent thermal conductivity of samples (a) irradiated by different Ga$^+$ ion energies and (b) with different crystallographic orientations of the β-Ga$_2$O$_3$ substrate implanted with 0.5 MeV Ga$^+$ ions in comparison with the pristine β-Ga$_2$O$_3$

Table I. Thermal conductivity $k$ of γ-Ga$_2$O$_3$ and interface thermal conductance ($G$) across the Al/γ-Ga$_2$O$_3$ interface for samples implanted with different ion energies based on three layers (Al/γ-Ga$_2$O$_3$/β-Ga$_2$O$_3$).

| Ion energy (MeV) | d of γ-Ga$_2$O$_3$ films (nm) | k of γ-Ga$_2$O$_3$ (W m$^{-1}$K$^{-1}$) | G of Al/γ-Ga$_2$O$_3$ (MW m$^{-2}$K$^{-1}$) |
|---|---|---|---|
| 0.5 | 350 | 1.84±0.21 | 48.43±5.51 |
| 1 | 650 | 1.87±0.23 | 51.73±6.19 |
| 1.7 | 1000 | 2.11±0.25 | 39.66±4.73 |

Further, using equilibrium molecular dynamics (EMD) simulations we calculated $k$ values of $β$-Ga$_2$O$_3$ for three different crystal directions, see the supplementary material. Notably, significant oscillations in the autocorrelation function while using the Green-Kubo relation inherently contribute to high uncertainties for $k$, because of the high noise levels preventing the identification of the convergence region. Therefore, twenty independent trajectories were chosen to improve the statistics, so that the conductivity reached the equilibration after 240 ps correlation time. These calculated $β$-Ga$_2$O$_3$ thermal conductivity results using the Born-Mayer-Huggins potential[31] are in good agreement with literature data[18,19,31] exhibiting a strong anisotropy with the largest conductivity along [010] direction (see Supplementary Materials). On the other hand, the tabGAP potential[32] underestimates the $k$ value for β-phase resulting in 11.3±0.7 W/m-K along [010] direction which is almost 2 times lower than the Born-Mayer-Huggins potential and experimental values reported before (see Supplementary Materials). Nevertheless, using the same approach and two different potentials, we performed calculations of the $k$ in the $γ$-phase, see the data in Fig. 3 (a). For these simulations, the saturation of the $k$-values were achieved at 160 ps, resulting in $k$ of 4 W m$^{-1}$ K$^{-1}$ for the Born-Mayer-Huggins potential and 3.1 W m$^{-1}$ K$^{-1}$ for the tabGAP potential, i.e. showing qualitatively consistent trend with the experimental values in Fig. 2 and Table I; however, exceeding the experimental values approximately by a factor of two.

For more insights into mechanisms, we performed complementary NEMD simulations using Born-Mayer-Huggins potential, specifically Figure 3 (b) shows the evolution of $1/k$ as a function of the inverse size of the simulation box in $γ$-Ga$_2$O$_3$. Figure 3 (b) shows an apparent linear trend, with the intercept of the $1/k$ axis giving the $k$ value for an "infinitely large" box corresponding to a bulk thermal conductivity of 3.82 W m$^{-1}$ K$^{-1}$ for γ-Ga$_2$O$_3$ consistently with EMD results.

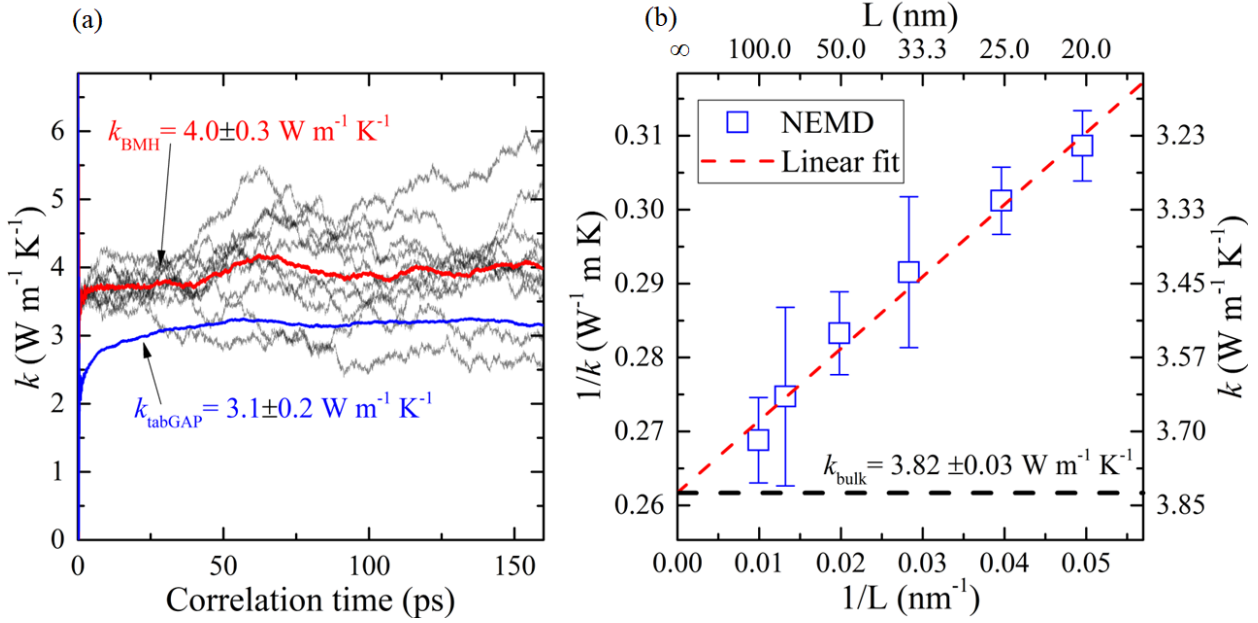

Fig. 3. Modeling of the heat conduction in $\gamma$-Ga$_2$O$_3$ using MD; (a) $k$ versus correlation time obtained via EMD for Born-Mayer-Huggins (BMH) and tabGAP potential and (b) $1/k$ as a function of the inverse size of the simulation box for the NEMD simulations using Born-Mayer-Huggins potential. In panel (b) the red dashed line corresponds to a linear fit to the NEMD data and used to extrapolate the bulk $k$ of $\gamma$-Ga$_2$O$_3$ (black dashed line).

At this end, Fig. 4 (a) summarizes $k$-values for $\beta$- and $\gamma$-Ga$_2$O$_3$ obtained in the present work in comparison with literature data for different forms of Ga$_2$O$_3$[21,22,30,33,34]. Most importantly, our experimental and simulation results show that the thermal conductivity of $\gamma$-phase is significantly lower than that of other Ga$_2$O$_3$ crystal phases, which can be attributed to a lower phonon mean free path $\lambda_{mfp} \approx 4$ nm in $\gamma$-phase as extracted from the NEMD simulations. Notably, the NEMD results show that in $k$-values of $\gamma$-phase are quite independent of the film thickness in comparison with that in $\beta$-Ga$_2$O$_3$ thin films[33]. This is comparable with the trend observed in $\beta$-(Al$_x$Ga$_{1-x}$)$_2$O$_3$ thin films with different Al contents, where the $k$ variations were mainly attributed to the phonon-alloy disorder scattering[34] implying that the thermal transport in these structures is dominated mostly by the vibrational modes with short $\lambda_{mfp}$. Notably, it was reported that $\beta$-(Al$_x$Ga$_{1-x}$)$_2$O$_3$ converts to its $\gamma$-phase when the Al concentration exceeds 40 %[35,36].

Importantly, as already mentioned above, our experimentally and theoretically assessed $k$-values in $\gamma$-Ga$_2$O$_3$ are in qualitative agreement; however, the experimental data in Fig. 2 are approximately twofold lower than the MD data. This difference may be attributed to the fact that the MD simulations in Fig. 3 have not accounted for a potential strain accumulation in the double $\gamma/\beta$-Ga$_2$O$_3$ samples as was discussed in the literature[9]. Thus, to check whether the strain effect may explain the discrepancy we repeated the EMD simulations applying in-plane ($\varepsilon_{xx}$) and out-of-plane ($\varepsilon_{yy}$) strain in ratios similar to that reported in the literature[9]. The corresponding simulations calculations show, see Supplementary Materials, that the $k$-values of the γ-phase exhibit a decreasing trend as a function of the applied strain reaching the experimentally measured $k$ value at $\varepsilon_{xx}$=6.3 % and $\varepsilon_{yy}$=1.86 %. Notably, even though the impact of strain, as discussed above, seems to be reasonable in the context of literature, at the present stage, we cannot rule out other reasons for the reduction of heat conductivity, e.g. associated with ion-induced point defects and/or the imperfections in the γ-films[37].

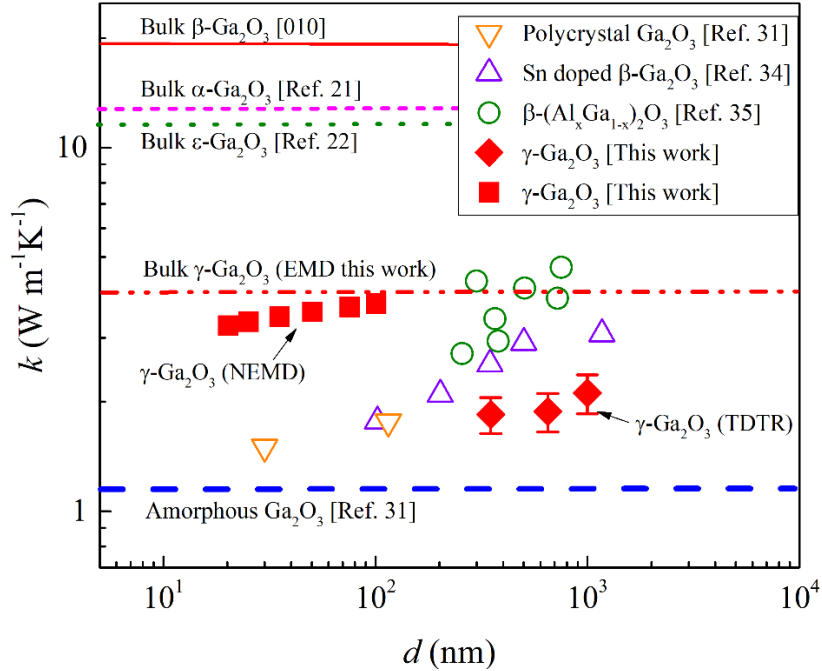

Fig. 4. Summary of the $k$ data measured in this work in comparison with the results available in literature[21,22,30,33,34]. Blue dashed line in panel (a) is the minimum $k$ of $Ga_2O_3$ obtained using Cahill's amorphous limit[29,30].

## IV. Conclusions

This work paves a new direction for studying nanoscale heat propagation in double *γ/β*-$Ga_2O_3$ polymorph structures, potentially applicable to modulate functional thermal properties across interfaces without changing the chemical composition of materials, in contrast to that in conventional heterostructures. Consistently with literature, the samples were fabricated by a self-organized *β*-to-*γ* polymorph transformation upon reaching a certain disorder threshold induced into *β*-$Ga_2O_3$ by the ion irradiation; as controlled by a combination of the XRD and RBS-C. To make a systematic comparison of the thermal properties, we depth-profiled the thermal conductivity across γ/β-Ga$_2$O$_3$ structures having variable γ-layer thicknesses (350 – 1000 nm) on top of the β-Ga$_2$O$_3$ substrates, using TDTR. As a result, in γ-Ga$_2$O$_3$, $k$ = 1.84 ÷ 2.11 W m$^{-1}$ K$^{-1}$ was measured independently of the initial β-substrates orientations. In parallel, we performed heat conduction MD simulations in the γ- and β-Ga$_2$O$_3$ lattices, using both equilibrium and non-equilibrium approaches. Importantly, for the relaxed γ-Ga$_2$O$_3$ lattice, in the MD simulations, we obtained approximately two-fold higher $k$ in comparison with the experimental results. Consistently with literature, such discrepancy may be attributed to the residual biaxial strain remaining in the γ-Ga$_2$O$_3$ lattice after the β-to-γ transition, even though other reasons for the reduction of thermal conductivity associated with the imperfections in the new γ-films could not be ruled out. In its turn, for monoclinic β-Ga$_2$O$_3$, the heat propagation showed a distinct anisotropy, with values ranging from 10 W m$^{-1}$ K$^{-1}$ for [201] to 20 W m$^{-1}$ K$^{-1}$ for [010] orientations, consistently with the MD simulations. Thus, these results demonstrate the viability of the nanoscale thermoreflectance depth-profiling metrology across chemically identical polymorph interfaces and open an avenue to explore the nanoscale phonon transport in such "polymorph heterostructures" exhibiting variations in thermal conductivities, e.g., of an order of magnitude in γ/β Ga$_2$O$_3$ formed on [010] oriented β-substrates.

**SUPPLEMENTARY MATERIALS**

The details of data analysis and molecular dynamics simulations are given in the Supplementary Materials.

**ACKNOWLEDGMENTS**

This work was supported by the Science Committee of the Ministry of Science and Higher Education of the Republic of Kazakhstan [Grants AP19577063, AP19679332], Nazarbayev University grants via Collaborative Research Program (CRP) [11022021CRP1504], Faculty Development Competitive Research Grants Program (FDCRGP) [20122022FD4130], and M-ERA.NET GOFIB project [Research Council of Norway project number 337627]. The international collaboration was in part enabled by the INTPART Program funded by the Research Council of Norway, [project number 322382].

**AUTHOR CONTRIBUTIONS**

**Azat Abdullaev:** Conceptualization, Methodology, Validation, Investigation, Writing-Original Draft, Writing-Review and Editing, Visualization. **Kairolla Sekerbayev:** Methodology, Software, Formal analysis, Investigation. **Alexander Azarov:** Validation, Formal analysis, Investigation. **Vishnukanthan Venkatachalapathy:** Validation, Formal analysis, Investigation. **Vinay S. Chauhan:** Validation, Formal analysis, Investigation. **Zhandos Utegulov:** Conceptualization, Methodology, Investigation, Writing-Review and Editing, Supervision, Funding acquisition. **Andrej Kuznetsov:** Conceptualization, Methodology, Resources, Writing-Review and Editing, Supervision, Funding acquisition.

## AUTHOR DECLARATIONS

The authors declare no conflict of interest.

## DATA AVAILABILITY

The data that support the findings of this study are available within the article and its supplementary material.

# Supplementary Materials

# Thermal Conductivity of Double Polymorph $Ga_2O_3$ Structures

Azat Abdullaev[1,2*], Kairolla Sekerbayev[1,2], Alexander Azarov[3], Vishnukanthan Venkatachalapathy[3], Vinay S. Chauhan[4], Zhandos Utegulov[2*], Andrej Kuznetsov[3*]

[1]*Center for Energy and Advanced Materials, Science National Laboratory of Astana, Nazarbayev University, Kabanbay Batyr avenue 53, 010000 Astana, Kazakhstan*
[2]*Department of Physics, School of Sciences and Humanities, Nazarbayev University, Kabanbay Batyr avenue 53, 010000 Astana, Kazakhstan*
[3]*Department of Physics, Centre for Materials Science and Nanotechnology, University of Oslo, N-0316 Oslo, Norway*
[4]*Intel Corporation, Chandler, 85226, Arizona, USA*

Corresponding authors: azat.abdullaev@nu.edu.kz, zhutegulov@nu.edu.kz, andrej.kuznetsov@fys.uio.no

*Section S1. Thermal conductivity of pristine β-$Ga_2O_3$.*

Fig. S1 (a) shows a ratio of the thermorelfectance signal for Al/β-$Ga_2O_3$ pristine sample. The inset shows a zoomed view of the first 80 ps. An acoustic echo from the Al/β-$Ga_2O_3$ interface is seen around 26 ps, which corresponds to Al thickness of 83 nm assuming the sound velocity of Al to be 6420 m/s ($d = \vartheta_{Al} * t/2$). In addition, an oxide layer of around 3 nm is also taken into account resulting in the total thickness of Al to be 86 nm. This result was also validated by a profilometer which measured the thickness of 85÷90 nm. The specific heat of Al and is taken from the literature[1]. Fig. S1 (b) is the resulted fitting of the experimental data at a single frequency employing a heat diffusion model across Al/β-$Ga_2O_3$ as mentioned in the main text. Table SI lists the extracted *k* for different orientations of β-$Ga_2O_3$ derived by simultaneously fitting several frequencies that are in agreement with literature values[2–4]. The error bars are set based on the Levenberg-Marquardt algorithm in accordance with literature [5,6].

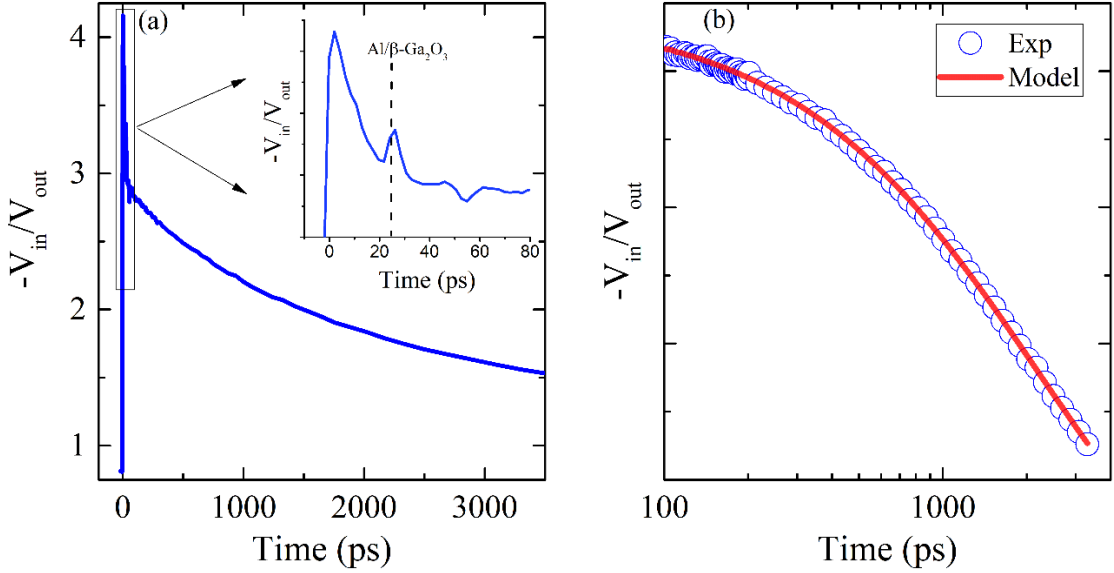

Fig. S1. (a) Transient thermoreflectance signal of pristine $\beta$-Ga$_2$O$_3$ covered with a thin Al layer. The inset zooms the first 80 ps, including the acoustic echo at the interface between Al and $\beta$-Ga$_2$O$_3$. (b) Experimental data and corresponding fitting obtained by the heat diffusion model for a single frequency.

Table SI. Extracted thermal conductivity of pristine β-Ga$_2$O$_3$ for different crystal orientations using multi-frequency fitting.

| Crystal orientation | [010] | [100] | [201] |
|---|---|---|---|
| $k$ (W m$^{-1}$ K$^{-1}$) | 19.3±2.05 | 13.81±1.52 | 11.75±1.51 |

*Section S2. TDTR fitting parameters*

Table SII lists the parameters that were used in the multi-frequency fitting of Al/$\gamma$-Ga$_2$O$_3$/$\beta$-Ga$_2$O$_3$ structure. The properties of Al and $\beta$-Ga$_2$O$_3$ are taken from literature as it is explained in Section S1 and in the main text. The thickness of $\gamma$-Ga$_2$O$_3$ was taken from the RBS-C measurements described in the main text. The G between $\gamma$-Ga$_2$O$_3$ and $\beta$-Ga$_2$O$_3$ was fixed at 250 MW m$^{-2}$K$^{-1}$. The fitted parameters are $k$ of γ-Ga$_2$O$_3$ and $G$ between Al and γ-Ga$_2$O$_3$. There is no information in literature about the specific heat capacity ($C_v$) of γ-Ga$_2$O$_3$ since this crystal phase was not observed in bulk and only recently in the thin film form. Therefore, $C_v$ of both γ- and β-phases were calculated using ALAMODE package[7] and soapGAP machine-learning IAP[8]. The results, as shown in Fig. S2, suggest that both polymorphs have close heat capacity values.

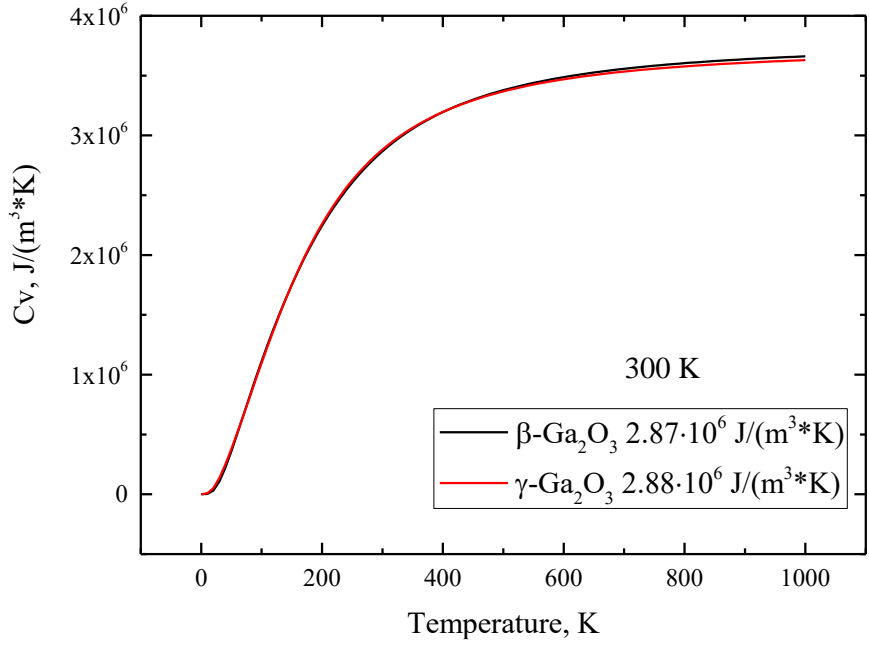

Fig. S2. Calculated volumetric specific heat capacity $C_v$ of $\gamma$- and $\beta$-phases at room temperature using ALAMODE package.

Table SII. Parameters used in the TDTR data fitting at room temperature.

| | $k$ (W m$^{-1}$ K$^{-1}$) | $C_V$ (J/ m³-K) | Thickness (µm) |
|---|---|---|---|
| Al | 160 | 2.42e6 | 0.086 |
| Interface | Fit | 0.1 | 0.001 |
| $\gamma$-Ga$_2$O$_3$ (0.5 MeV, 1 MeV, 1.7 MeV) | Fit | 2.87e6 | (0.35, 0.65, 1) |
| Interface | 2 | 2.8e6 | 0.008 |
| $\beta$-Ga$_2$O$_3$ | 20 | 2.88e6 | 1000 |

*Section S3. Parameters to model Al/$\gamma$-Ga$_2$O$_3$/$\beta$-Ga$_2$O$_3$ structures.*

Fig. S2 shows the results of the sensitivity analysis which is defined by the following expression[9]:

$$S_i = \frac{\partial \ln(\frac{-V_{in}}{V_{out}})}{\partial \ln(p_i)},$$

where $p_i$ is the value of parameter $i$ and $\frac{-V_{in}}{V_{out}}$ is the TDTR signal. Also the sensitivity analysis was used to determine the uncertainty of $k$. Typically, in the measurements the uncertainty in Al thickness was ~5%, $\gamma$-Ga$_2$O$_3$ thickness ~5%, volumetric specific heat capacity of $\gamma$- and $\beta$-Ga$_2$O$_3$

~5%, and spot size ~3%. Higher sensitivity leads to more accurate measurements. The results of the sensitivity analysis for Al/$\gamma$-Ga$_2$O$_3$/$\beta$-Ga$_2$O$_3$ structure are given in Fig. S3 and the schematics of this three-layer structure with corresponding heat penetration depth is also given in Fig. S4, illustrating 350 nm of $\gamma$-Ga$_2$O$_3$. As it is seen from Fig. S3, in the frequency range between 1.9 to 9.8 MHz, our measurements are strongly sensitive to the cross-plane conductivity ($k_z$) of the $\gamma$-phase and interface boundary conductance ($G$) between Al and $\gamma$-Ga$_2$O$_3$. However, for 0.73 MHz, the sensitivity of the β-phase increases dramatically, while that for the $\gamma$-phase drops. From this, we conclude that for the 350 nm thick sample, our measurements are valid at high frequencies, allowing the direct extraction the $\gamma$-phase conductivity.

Notably, in Fig. 2 (a) of the main text, the $k$ values for this sample (open square) at 1.9 MHz do not reach the pristine $\beta$-Ga$_2$O$_3$ value (filled square). This is because the heat penetration depth at 1.9 MHz is close to the interface region between $\gamma$ and $\beta$ phases, as such resulting into the mixed value.

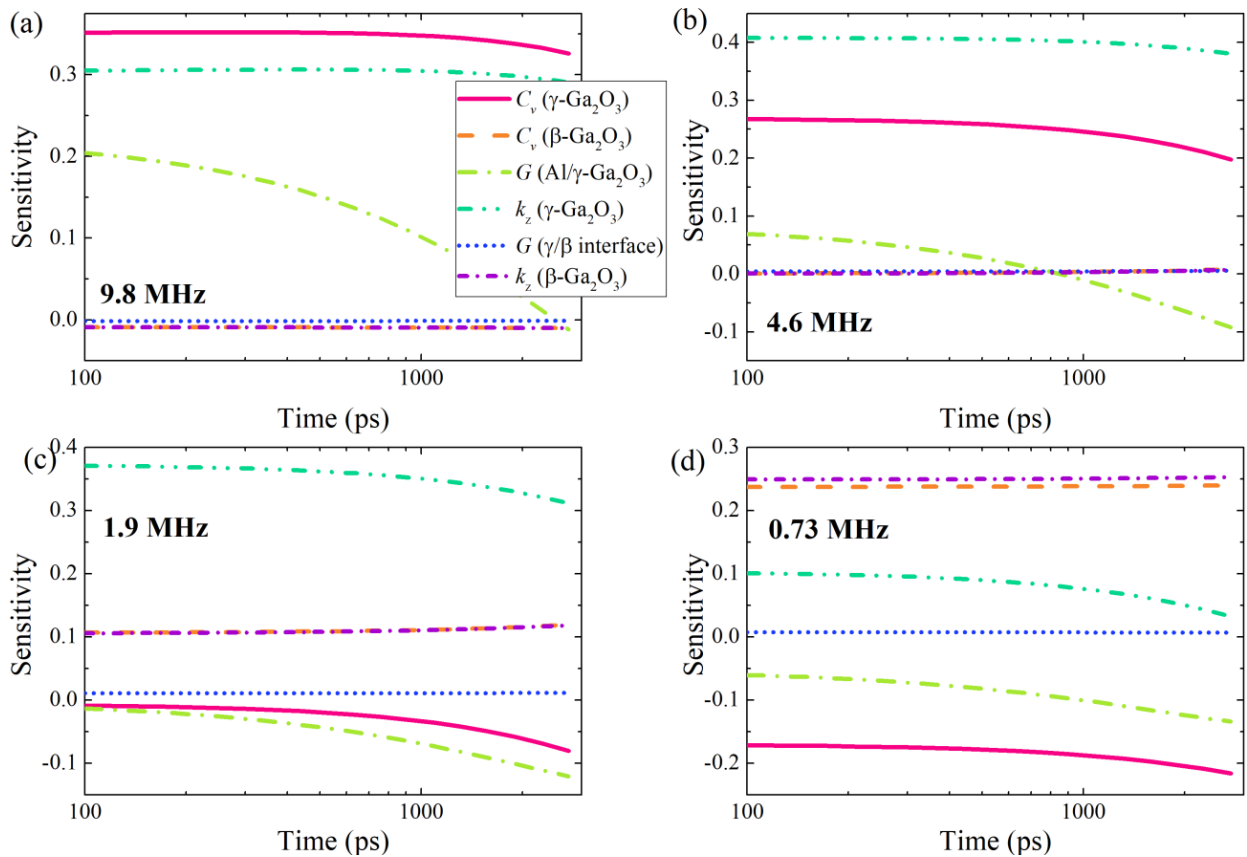

Fig. S3. Sensitivity analysis of $k_z$ and $G$ as a function of time delay for Al/$\gamma$-Ga$_2$O$_3$/$\beta$-Ga$_2$O$_3$ structure with the thickness of $\gamma$-Ga$_2$O$_3$ as 350 nm for different modulation frequencies: (a) 9.8 MHz, (b) 4.6 MHz, (c) 1.9 MHz and (d) 0.73 MHz.

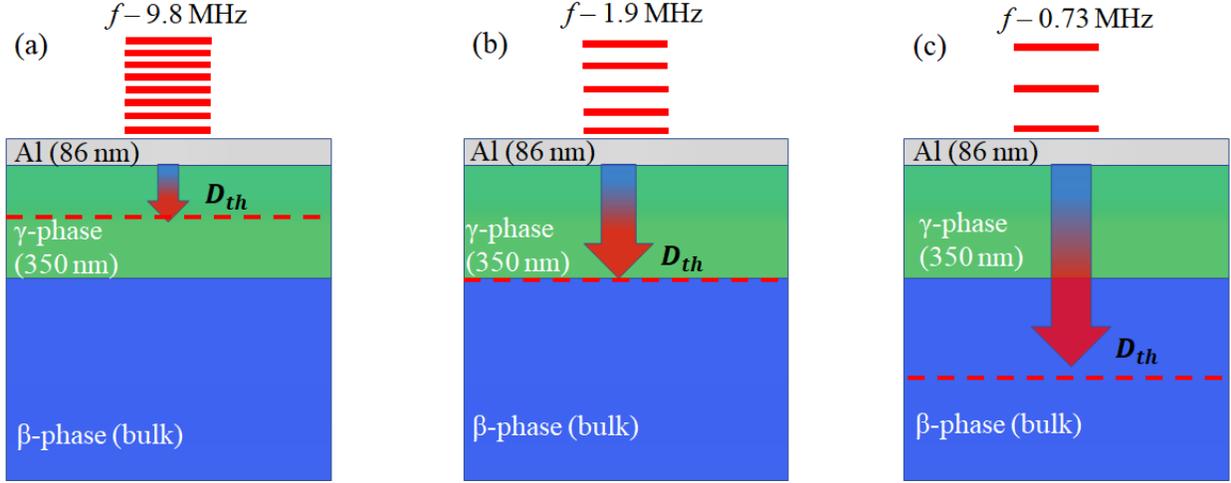

Fig. S4. Schematics of a three-layer structure (Al/γ-Ga$_2$O$_3$/β-Ga$_2$O$_3$) for 3 different modulation frequencies (a) 9.8 MHz, (b) 1.9 MHz and (c) 0.73 MHz. $D_{th}$ is a heat penetration depth which is inversely proportional to $f$.

*Section S4. EMD and NEMD calculations.*

In the EMD calculations, the lattice $k$ was modeled with the Green-Kubo formalism relating the conductivity to the heat current autocorrelation function:

$$k_x = \frac{V}{k_B T^2} \int_0^t \langle J_x(t) J_x(0) \rangle d\tau, \qquad (1)$$

where $k_x$ – is one dimensional thermal conductivity, $V$ is the volume of the simulation cell, $k_B$ is the Boltzmann constant, $T$ is the absolute temperature and $t$ is the correlation time. The heat flux $J_x$ is defined as:

$$J_x = \frac{1}{V}\left[\sum_i e_i \boldsymbol{v}_i - \sum_i \boldsymbol{S}_i \boldsymbol{v}_i\right], \qquad (2)$$

where $e_i$ is the total energy, and $\boldsymbol{v}_i$ is the velocity of the $i_{th}$ atom, $\boldsymbol{S}_i$ is the per-atom stress tensor.

The EMD were employed to β-Ga$_2$O$_3$ with a supercell size of 4.9 × 4.1 × 4.1 nm$^3$ containing 7280 atoms and to γ-Ga$_2$O$_3$ with a supercell size of 4.2 × 4.2 × 5.0 nm$^3$ containing 8000 atoms. Before performing the Green-Kubo integration the supercell was equilibrated for 250 ps with the isobaric-isothermal ensemble (NPT) at 300 K and ambient pressure (1 bar) with the subsequent canonical ensemble (NVT) held at 300 K for 400 ps. The time step was 2 fs and the correlation time was 160–240 ps with a total accumulation duration of 3.2–4.8 ns. We used Born–Mayer–Huggins type expression[10] for simulating the pairwise atom interaction both for β- and γ-Ga$_2$O$_3$.

This type of interatomic potential is conventionally used to estimate thermal properties of oxide systems, including $\beta$-Ga$_2$O$_3$[10]. We have used the same parameters as in[10] employed for $\beta$-Ga$_2$O$_3$. Notably, Born–Mayer–Huggins type of potential was originally developed for $\beta$-phase. To cross-validate the simulation results, a machine-learning-based potential (the tabGAP) trained for different polymorphs and disordered Ga$_2$O$_3$ structures was used as well[8]. The simulation parameters for the tabGAP potential were the same as for the Born-Mayer-Huggins potential.

Fig. S5 and Fig. S6 show the EMD calculation of $k$ for $\beta$-Ga$_2$O$_3$ in different crystallographic directions using Born-Mayer-Huggins potential and tabGAP potential, respectively. The results for Born-Mayer-Huggins potential are in a good agreement with the previous works[10], while tabGAP potential underestimates the conductivity along [010] direction.

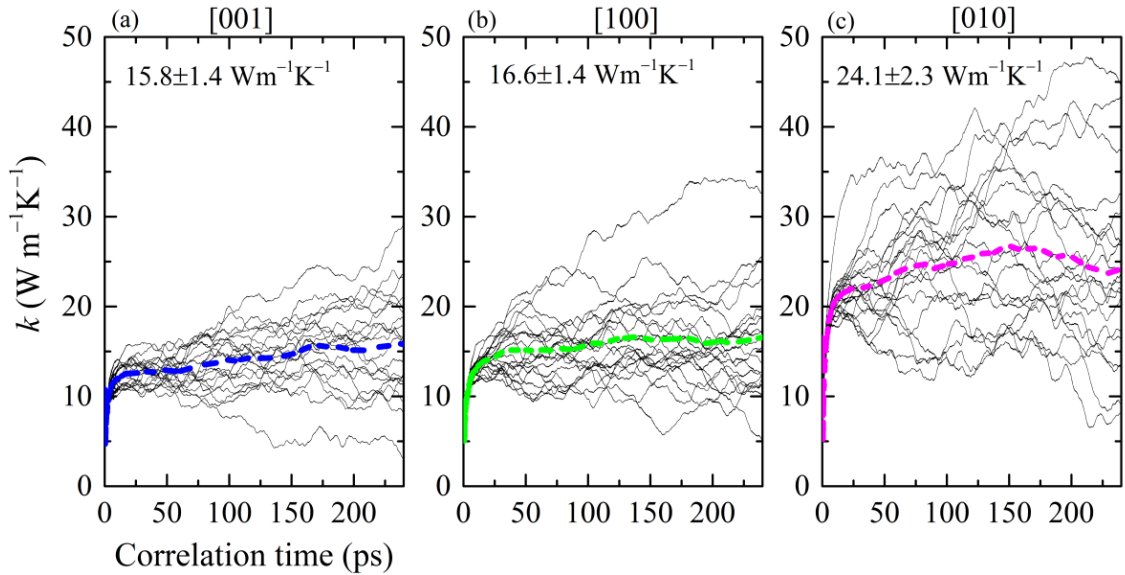

Fig. S5. $k$ of $\beta$-Ga$_2$O$_3$ versus correlation time obtained via EMD using Born-Mayer-Huggins potential.

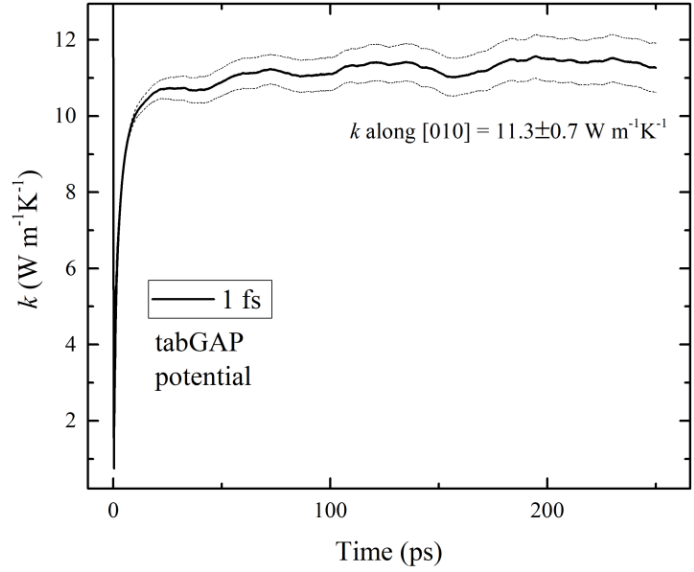

Fig. S6. Calculated $k$ of $\beta$-Ga$_2$O$_3$ in [010] direction using tabGAP potential.

A defined volume is spanned with repeating unit cells of $\gamma$-Ga$_2$O$_3$. First, the system reaches the equilibration at 300 K with a time step of 2 fs over 100 000 time steps followed by further 600 000 time steps, simulating the effects of the heat source/sinks, so that a non-equilibrium steady state is reached. Finally, temperature profile data are averaged over additional 200 000 time steps. Linear region of the temperature change from the cold to hot reservoirs is fitted to extract temperature gradient $\Delta L/\Delta T$ as depicted in Fig. S7. We extract $k$ from temperature gradients using Fourier heat conduction law $k = \frac{Q \Delta L}{S \Delta T}$, where $Q$ is the amount of heat transferred across an area $S$ per unit time. A linear relationship between the inverse $k$ and the inverse size of the simulation cell is used to extract the bulk $k$[11]:

$$\frac{1}{k(L)} = A + \frac{B}{L}, \qquad (3)$$

where $L$ is the simulation box length, $k(L)$ is heat conductivity at the corresponding length $L$, while $A$ and $B$ are the fitting parameters. Thus, $k_{bulk} = 1/A$ and $\lambda_{mfp} = B/A$.

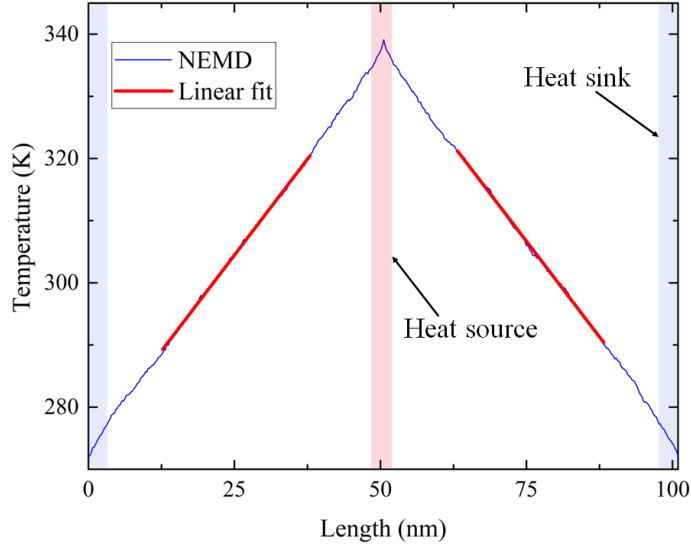

Fig. S7. Temperature profile along c-axis which corresponds to [100] direction.

Figure S8 shows the EMD calculated thermal conductivity of γ-phase with and without strain as a funciton of correlation time using Born-Mayer-Hugging (BMH) and tabGAP potentials, while Figure S9 shows the *k*-values as a function of applied strain for BMH potential only.

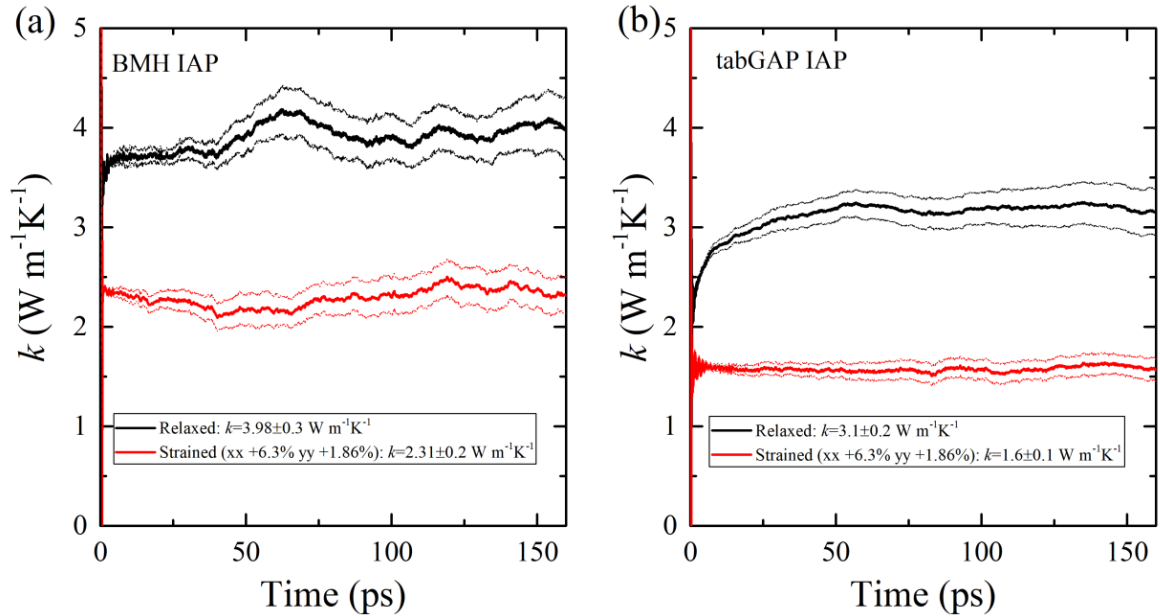

Fig. S8. *k* versus correlation time obtained via EMD for (a) Born-Mayer-Huggins (BMH) and (b) tabGAP potential with and without applied strain.

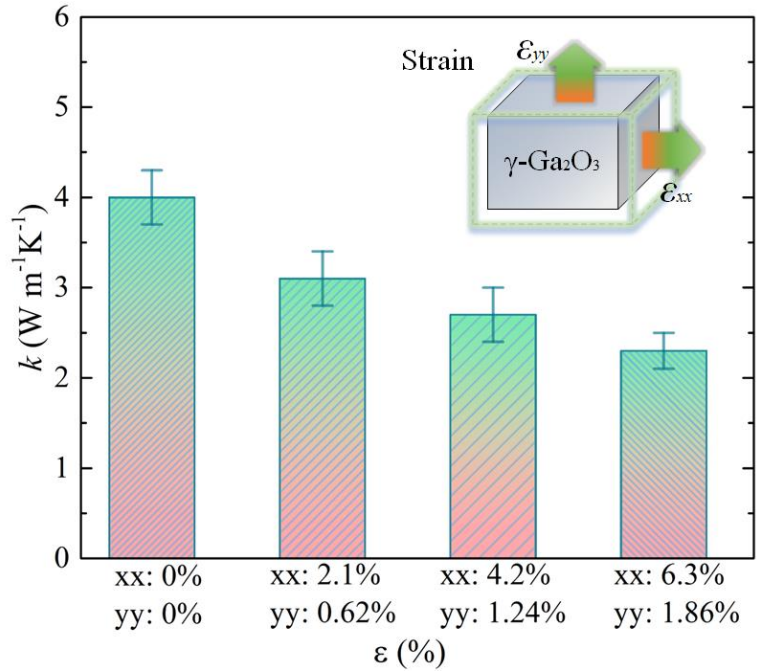

Fig. S9. Illustration of the γ-Ga$_2$O$_3$ $k$ variations as a function of strain applied in the EMD simulations using BMH potential.